\documentclass[11pt]{article}
\usepackage{blois,epsfig}

\def\Journal#1#2#3#4{{#1} {\bf #2}, #3 (#4)}

\usepackage{amsmath}
\usepackage{graphicx}
\usepackage{multirow}
\usepackage{epsfig}
\usepackage{rotating}

\def\NPB{{\em Nucl. Phys.} B}
\def\PLB{{\em Phys. Lett.}  B}
\def\PRL{\em Phys. Rev. Lett.}
\def\PRD{{\em Phys. Rev.} D}

\def\be{\begin{equation}}
\def\ee{\end{equation}}
\def\beq{\begin{eqnarray}}
\def\eeq{\end{eqnarray}}

\begin{document}
\vspace*{4cm}
\title{$Z'$ SEARCHES: FROM TEVATRON TO LHC}

\author{J. ERLER$^a$, P. LANGACKER$^b$, S. MUNIR$^a$, E. ROJAS$^a$}

\address{$^a$Departamento de F\'isica Te\'orica, Instituto de F\'isica, \\ 
Universidad Nacional Aut\'onoma de M\'exico, 04510 M\'exico D.F., M\'exico \\
$^b$School of Natural Sciences, Institute for Advanced Study, \\ 
Einstein Drive, Princeton, NJ 08540, USA}

\maketitle

\abstracts{The CDF collaboration has set lower limits on the masses of the $Z'$ bosons occurring in a range of $E_6$ GUT based models. We revisit their analysis and extend it to certain other $E_6$ scenarios as well as to some general classes of models satisfying the anomaly cancellation conditions, which are not included in the CDF analysis. We also suggest a Bayesian statistical method for finding exclusion limits on the $Z'$ mass, which allows one to explore a wide range of the $U(1)'$ gauge coupling parameter. This method also takes into account the effects of interference between the $Z'$ and the SM gauge bosons.}

\section{Introduction}
A neutral $Z'$ gauge boson appears in numerous models containing the SM gauge symmetry group along with an additional $U(1)$ symmetry (for a review, see~\cite{Langacker:2008yv}). Grand Unified Theories (GUTs) larger than the original $SU(5)$ model, such as $SO(10)$~\cite{Robinett:1982tq} or $E_6$~\cite{Langacker:1984dc,Hewett:1988xc}, break down to the SM as: $E_6 \rightarrow SO(10)\times U(1)_\psi$, with in turn $SO(10)\rightarrow~SU(5)~\times~U(1)_\chi~\rightarrow$ SM $\times~U(1)_\chi$~\cite{Georgi:1974sy}. The $Z'$ thus surviving at the electroweak (EW) scale can be written as the linear combination,
\beq
\label{eq:e6kinmix}
Z' = \cos\alpha \cos\beta Z_\chi + \sin\alpha \cos\beta Z_Y + \sin\beta Z_\psi.
\eeq

If kinetic mixing~\cite{Holdom:1985ag,delAguila:1995rb,Babu:1996vt} with the hypercharge group $U(1)_Y$ is neglected by setting $\alpha=0$ in eq. (\ref{eq:e6kinmix}), one obtains some well--known $Z'$ bosons by adjusting $\beta$. These include $Z_\chi$ ($\beta = 0^\circ$), $Z_\psi$ ($\beta = 90^\circ$), $Z_\eta$ ($\beta \approx -52.2^\circ$)~\cite{Candelas:1985en}, $Z_I$ ($\beta \approx 37.8^\circ$)~\cite{Langacker:1984dc}, $Z_S$ ($\beta \approx 23.3^\circ$)~\cite{Erler:2002pr,Kang:2004pp} and $Z_N$ ($\beta \approx 75.5^\circ$)~\cite{Ma:1995xk}. The inclusion of kinetic mixing results in certain other phenomenologically interesting cases, such as $Z_{dph}$ $(\alpha, \beta \approx -78.5^\circ, 37.8^\circ)$ which does not couple to the $d$--type quarks, $Z_R$ ($\alpha, \beta \approx 50.8^\circ,0^\circ$)~\cite{Erler:2009jh} which couples to the right--handed fermions only and $Z_{B-L}$ $(\alpha,\beta \approx - 39.2^\circ, 0^\circ$)~\cite{Appelquist:2002mw}, where $B$ is the baryon number and $L$ the lepton number of an ordinary fermion. The $Z_{LR}$ which exists in models with left--right symmetry~\cite{Mohapatra:1986uf} is equivalent to the linear combination: $\sqrt{3/5}\, (\bar\alpha\, Z_R - Z_{B-L}/2\bar\alpha)$, where $\bar\alpha \equiv \sqrt{g_R^2/g_L^2 \cot^2 \theta_W - 1}$, with $\theta_W$ being the weak mixing angle and $g_{L,R}$ being the $SU(2)_{L,R}$ coupling strengths, respectively. The $Z_{LR}$ studied here corresponds to the specific case of $g_L=g_R$. In addition to these $E_6$ based models, a $Z_{string}$ from a specific superstring model~\cite{Chaudhuri:1994cd} and a sequential $Z_{SM}$ are also included in this analysis.

A number of other classes of one or more--parameter models have been discussed \cite{Erler:2000wu,Carena:2004xs,Langacker:2008yv}. Without the assumption of unification at the GUT scale, but assuming nullification of anomalies using three families of exotics (which is not the case in some supersymmetric models), there are four `one-parameter' models \cite{Carena:2004xs}, with fermion charges $B-xL$, $d-xu$, $q+xu$ and $10+x\bar{5}$, with $x$ arbitrary. The last three of these correspond to the generalized $E_6$ charges in eq. (\ref{eq:e6kinmix}). ($q+xu$ corresponds to arbitrary superpositions of $Y$ and $B-L$, while $10 + x \bar{5}$ corresponds to the $E_6$ models without kinetic mixing.) We have, therefore, normalized the fermions charges, $z_f$, in the $x$--parameter models to their $E_6$ values, $\epsilon_f$. These charges are given in Table~\ref{tab:x-charges}.  
\begin{table}[t]
\caption{$E_6$ fermion charges, $\epsilon_f$, in terms of $z_f$, their values in the corresponding $x$--parameter models. Also given are the angles $\alpha$ and $\beta$ yielding these models.\label{tab:x-charges}}
\vspace{0.4cm}
\begin{center}
\begin{tabular}{|c|c|c|c|}
\hline 
            &$q+xu$&$10+x\bar{5}$&$d-xu$\\  \hline
$\epsilon_f$&$\frac{3}{2\sqrt{7-2x+x^2}} z_f$
            &$-\frac{3}{2\sqrt{4+x+x^2}} z_f$&$\frac{3}{2\sqrt{1-x+x^2}} z_f$ \\
            \hline
$\tan \beta$& $0$  & $-\sqrt{\frac{3}{5}}(\frac{3+x}{1-x})$ &
               $\frac{\sqrt{3}(1-x)sign(5-x)}{\sqrt{5-2x+5x^2}}$  \\   \hline
$\tan \alpha$&$\sqrt{\frac{3}{2}}(\frac{1+x}{x-4})$&$0$ &
              $2\sqrt{6}(\frac{x}{5-x})$ \\   \hline
\end{tabular}
\end{center}
\end{table}

\section{$Z'$ production at CDF and limits on its mass}
The total cross--section for the Drell--Yan (DY) process at a hadron collider, with a neutral gauge boson $B$ as the mediator and $\mu^{+}\mu^{-}$ as the outgoing particles, is given as~\cite{Ellis:1991qj}
\beq
\label{eq:lo3}
\sigma =\frac{2}{s}\int_0^{\sqrt{s}}MdM\sigma_{diff},
\eeq
where $M$ is the invariant mass of the muon pair, $\sqrt{s}$ is the center--of--mass (CM) energy, and
\beq
\label{eq:lo4}
\sigma_{diff}=\int_{M^2/s}^1 dx_{1}\frac{1}{x_1}
\sum_{q} \hat{\sigma}(M^2)\frac{K}{N_c}\Big\{f_{q}^{A}(x_{1},M^2)f_{\bar{q}}^{B}(x_{2},M^2)+f_{\bar{q}}^{A}(x_{1},M^2)f_{q}^{B}(x_{2},M^2)\Big\},
\eeq
where $N_c=3$ is the quark color factor. $x_1$ and $x_{2}(\equiv\frac{M^2}{x_1s})$ above are the momentum fractions of the ingoing partons having parton distribution functions (PDFs) $f_{q/\bar{q}}^{A/B}$ for hadrons $A$ and $B$. $\alpha_s$ is the strong coupling constant and $K=K_{C}+K_{E}$, with $K_{C}$ being the QCD $K$--factor and $K_{E}$ being the multiplicative factor due to QED corrections. $\hat{\sigma}(M^2)$ is equal to
\beq
\label{eq:hard}
\int_{-1}^{1}d\cos \theta^* \frac{1}{128\pi M^2}\Big[
\left(\lvert A_{LL}\rvert^2+\lvert A_{RR}\rvert^2\right)(1+\cos\theta^*)^2+\left(\lvert A_{LR}\rvert^2+\lvert A_{RL}\rvert^2\right)(1-\cos\theta^*)^2\Big],
\eeq
with $\theta^*$ being the polar angle defined in the CM frame and the individual amplitudes given as
\beq\label{eq:amplitud}
 A_{ij}= -Qe^2+\frac{M^2}{M^2-M_{Z}^2+iM_Z\Gamma_Z}C^{Z}_i(q)C^Z_j(l)
+\frac{M^2}{M^2-M_{Z'}^2+iM_{Z'}\Gamma_{Z'}}C^{Z'}_i(q)C^{Z'}_j(l),
\eeq
where $i,j$ are run over $L,R$. $Q$ and $e$ are the electric charges of the contributing quark and the muon, respectively. $C^{Z,Z'}_{L,R}(f)\equiv g_{1,2}\epsilon_{L,R}^{Z,Z'}(f)$, with $g_1=e/\sin\theta_w \cos\theta_w$ being the gauge coupling strength of the $Z$ boson, $g_2$ the $Z'$ coupling and $\epsilon^{Z,Z'}_{L,R}$ the EW and $U(1)'$ charges of the fermion $f$. $\Gamma_{Z,Z'}$ are the total decay widths of the $Z$ and $Z'$ bosons having masses $M_{Z,Z'}$.

\begin{figure}
\begin{center}
\includegraphics[scale=0.38]{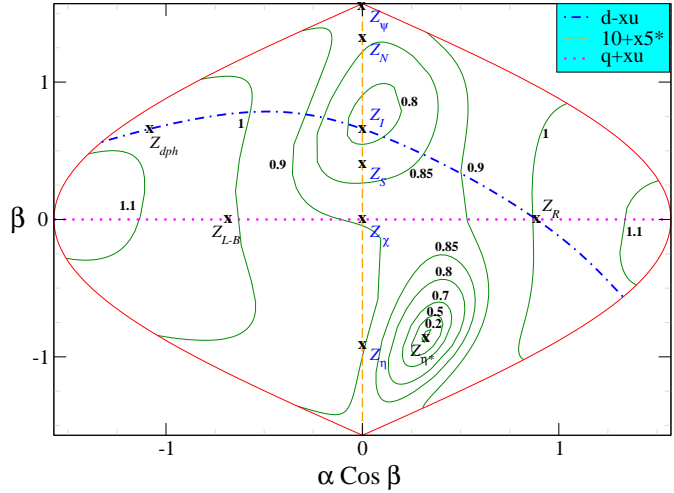}
\end{center}
\vspace{-0.3cm}
\caption{Contours in $M_{Z'}$ limits mentioned on/beside them. $x$ shows the location of a particular $E_6$ model within a contour. The dotted, dashed and dot--dashed lines correspond to the three $x$--parameter models. $Z_{\eta^*}$ is a leptophobic boson which will not be observable in the di--lepton channels.}
\label{fig:CDForig}
\end{figure}

We first follow the CDF analysis~\cite{Aaltonen:2008ah} and use the LO expression given in eq. (\ref{eq:lo3}) to calculate the cross--section due to the $Z'$ alone, neglecting the $\gamma$ and $Z$ contributions to the amplitudes in eq. (\ref{eq:amplitud}). We employ LO CTEQ6L PDFs \cite{Pumplin:2002vw} and mass--dependent NNLO $K_C$~\cite{Carena:2004xs} and NLO $K_E$~\cite{Baur:1997wa} values, and assume that the $Z'$ decays into SM fermions only, which are taken to be massless. We achieve up to 99.8\% agreement on the 95\% confidence level (C.L.) $M_{Z'}$ lower limits with the CDF analysis, for the $E_6$ models included therein, and obtain new limits for the rest of the models. The numerical values of the limits are given in Table~\ref{tab:limits} and the $\alpha,\beta$ parametrization of the models with contours in $M_{Z'}$ is plotted in Fig.~\ref{fig:CDForig}. In Table~\ref{tab:LHC} we give limits obtained for the LHC with a similar approach for some expected integrated luminosity and CM energy values.  

\begin{table}
\caption{Numerical limits in GeV on the mass of the $Z'$ boson in various models.}
\vspace{0.4cm}
\centering \begin{tabular}{|c|c|c|c|c|c|c|c|c|c|c|c|}
\hline
$Z_\chi$ & $Z_\psi$ & $Z_\eta$ & $Z_N$ & $Z_S$ & $Z_I$ & $Z_{B-L}$ & $Z_R$ & $Z_{LR}$ & $Z_{dph}$ & $Z_{string}$ &$Z_{SM}$\\\hline
895 & 883 & 910 & 865 & 823 & 790 & 1012 & 1006 & 959 & 1079 & 710 & 1030 \\\hline
\end{tabular}
\label{tab:limits}
\end{table}

\begin{table}[t!]
\caption{Limits on $M_{Z'}$ in the $Z_\chi$ model from the LHC.} 
\vspace{0.4cm}

\centering \begin{tabular}{|c|c|c|c|c|}
\hline
$\mathcal{L}$ (fb$^{-1}$)/$\sqrt{s}$ (TeV)&3.5&7&14&28\\\hline
3&1.0&1.85&3.0&4.65\\\hline
30&1.5&2.5&4.1&6.7\\\hline
300&2.2&3.3&5.4&9.7\\\hline
3000&3.9&5.55&7.9&12.2\\\hline
\end{tabular}
\label{tab:LHC}
\end{table}

\section{Bayesian statistical method}
The CDF analysis uses signal templates generated with a fixed resonance pole width, $\Gamma = 2.8\% \times M_{Z'}$. However, there is no fundamental reason to only look for such a narrow $Z'$. A wide $Z'$ resonance, implying a strongly coupling boson, could well be scattered over a few bins and no significant enhancement above the background will be visible. Besides, the effects of interference between the various bosonic contributions to the propagator, i.e., between $\gamma$, $Z$ and $Z'$ (see eq. (\ref{eq:amplitud})), are lost in their approach. These effects could in principle cause a considerable enhancement or dip in the number of events in several accompanying low $M^{-1}$ bins, e.g., in the case of a strongly interacting $Z'$ boson with mass just beyond the kinematic reach of the CDF. Finally, the CDF limits assume a fixed GUT--based $g_2$ and it is not straightforward to extend the limits to other $g_2$ values, particularly in the strong coupling regime (see, however~\cite{CDF2010}).

\begin{figure}
\begin{center}
\includegraphics[scale=0.37]{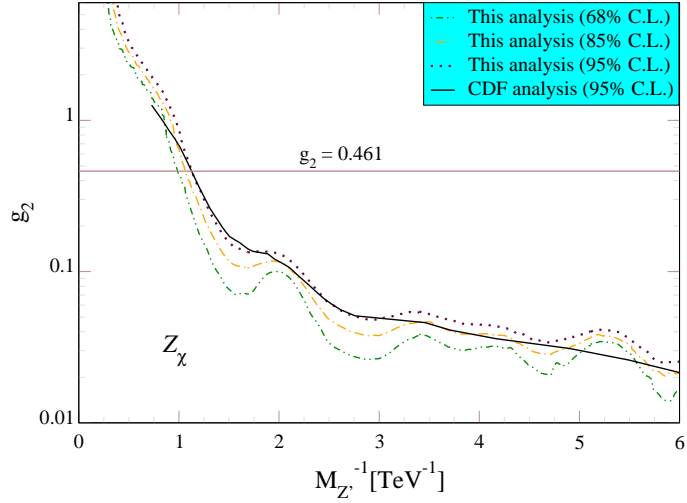}
\end{center}
\vspace{-0.3cm}
\caption{Exclusion contours for the $Z_\chi$ boson. The dotted, dot--dashed and double--dot--dashed curves correspond to the C.L. values given and the solid curve represents the CDF limits generalized to other $g_2$ values.}
\label{fig:CDFgrid}
\end{figure}

Therefore, we propose a Bayesian statistical method which allows one to vary $g_2$ in order to obtain the corresponding limit on $M_{Z'}$. It is based on the likelihood function $\L$, written as
\beq
\label{eq:likelihood}
\L(\vec{\mu}\lvert \vec{n})=\Pi_i^B P(n_i\lvert \mu_i),
\eeq
where $P$ is the Poisson probability of finding $n_i$ events given $\mu_i$ expected events in the $i$th bin with $B$ total bins. $\L$ in eq. (\ref{eq:likelihood}) is then evaluated using two hypotheses: the null hypothesis, $\L(\vec{\mu^b}\lvert \vec{n})$, assumes that the DY process occurs only via $\gamma$ and $Z$, and the signal hypothesis, $\L(\vec{\mu^t}\lvert \vec{n})$, with $\vec{\mu^t}=\vec{\mu^b}+\vec{\mu'}$, wherein the $Z'$ boson also contributes to the cross--section along with the SM gauge bosons. The SM events, $\mu^b$, expected in an invariant mass bin are calculated as 
\beq
\label{eq:nbin}
\mu^b = \mathcal{L}\sigma_{SM} = \frac{2E\mathcal{L}}{s} \int_{bin} AdM_p^{-1} \int^{\sqrt{s}}_0 MdM \textbf{p}(M^{-1}_p|M^{-1})\sigma_{diff} + n_{\text{BG}}, 
\eeq 
where $\mathcal{L}=2.3$ fb$^{-1}$ is the integrated luminosity at the CDF, $E=0.982$ is the detector {\it efficiency} and $A$ is the CDF {\it acceptance}, which is a mass--dependent multiplicative factor. $M_p$ in the above equation is the muon--pair mass measured by the detector, $n_{\text{BG}}$ refers to the non--DY events and the probability density is given as
\beq
\textbf{p}(M^{-1}_p|M^{-1}) = M_pb^a e^{-b}/\Gamma(a),
\eeq
with $a=(M^{-1}/\Delta)^2$, $b=M^{-1}M_p^{-1}/\Delta^2$, where $\Delta = 0.17$ TeV$^{-1}$ is the variance. The purpose of the above probability function is to smear over the DY background before distributing it into bins, hence accounting for the mis--identification of an event in a bin where it does not actually belong. For $\mu'$, a $\chi^2$ function is constructed as 
\beq
\chi_{\mu'}^2 = -2LLR = -2\ln(\frac{\L(\vec{\mu^t}\lvert \vec{n})}{\L(\vec{\mu^b}\lvert \vec{n})})= -2\sum_i^B(\mu_i^b-\mu_i^t + n_i\ln(\frac{\mu_i^t}{\mu_i^b}),
\eeq
and is minimized to obtain the best--fit values in the given range of $g_2$ and $M_{Z'}$, which correspond to a $Z'$ boson with cross--section $\mu'$ best favored by the data. The contours in $g_2$ and $M_{Z'}$, for a certain C.L. value specified by the allowed number of standard deviations, $\Delta\chi^2$, from the minimum, can then also be drawn, giving the exclusion limits on these parameters. Our preliminary contours are given in Fig. \ref{fig:CDFgrid} for $Z_{\chi}$ as a representative model. The numerical value of the 95\% C.L. limit is 913 GeV for $g_2=0.461$, which is about 21 GeV higher than the CDF value.
We next plan to undertake a global analysis including constraints from electroweak precision data~\cite{Erler:2009jh}. Eventually, a similar statistical analysis of the LHC data, as soon as it is released, will also be performed. 

\section*{Acknowledgments}
The work at IF-UNAM is supported by CONACyT project 82291--F. The work of P.L.~is supported by an IBM Einstein Fellowship and by NSF grant PHY--0969448.

\end{document}